\documentclass[prb,twocolumn,superscriptaddress,preprintnumbers,amsmath,amssymb]{revtex4}
\usepackage{graphicx}
\usepackage{amsmath}
\usepackage{latexsym}
\usepackage{dcolumn}
\usepackage{bm}
\usepackage{float}
\usepackage{graphicx,here}

\emergencystretch=\maxdimen
\begin{document}
\title{Experimental evidence of monolayer AlB$_2$ with symmetry-protected Dirac cones}

\author{Daiyu Geng}
\affiliation{Institute of Physics, Chinese Academy of Sciences, Beijing 100190, China}
\affiliation{School of Physical Sciences, University of Chinese Academy of Sciences, Beijing 100049, China}
\author{Kejun Yu}
\affiliation{Key Lab of Advanced Optoelectronic Quantum Architecture and Measurement (MOE), School of Physics, Beijing Institute of Technology, Beijing 100081, China}
\author{Shaosheng Yue}
\affiliation{Institute of Physics, Chinese Academy of Sciences, Beijing 100190, China}
\affiliation{School of Physical Sciences, University of Chinese Academy of Sciences, Beijing 100049, China}
\author{Jin Cao}
\affiliation{Key Lab of Advanced Optoelectronic Quantum Architecture and Measurement (MOE), School of Physics, Beijing Institute of Technology, Beijing 100081, China}
\author{Wenbin Li}
\affiliation{Institute of Physics, Chinese Academy of Sciences, Beijing 100190, China}
\affiliation{School of Physical Sciences, University of Chinese Academy of Sciences, Beijing 100049, China}
\author{Dashuai Ma}
\affiliation{Key Lab of Advanced Optoelectronic Quantum Architecture and Measurement (MOE), School of Physics, Beijing Institute of Technology, Beijing 100081, China}
\author{Chaoxi Cui}
\affiliation{Key Lab of Advanced Optoelectronic Quantum Architecture and Measurement (MOE), School of Physics, Beijing Institute of Technology, Beijing 100081, China}
\author{Masashi Arita}
\affiliation{Hiroshima Synchrotron Radiation Center, Hiroshima University, 2-313 Kagamiyama, Higashi-Hiroshima 739-0046, Japan}
\author{Shiv Kumar}
\affiliation{Hiroshima Synchrotron Radiation Center, Hiroshima University, 2-313 Kagamiyama, Higashi-Hiroshima 739-0046, Japan}
\author{Eike F. Schwier}
\affiliation{Hiroshima Synchrotron Radiation Center, Hiroshima University, 2-313 Kagamiyama, Higashi-Hiroshima 739-0046, Japan}
\author{Kenya Shimada}
\affiliation{Hiroshima Synchrotron Radiation Center, Hiroshima University, 2-313 Kagamiyama, Higashi-Hiroshima 739-0046, Japan}
\author{Peng Cheng}
\affiliation{Institute of Physics, Chinese Academy of Sciences, Beijing 100190, China}
\affiliation{School of Physical Sciences, University of Chinese Academy of Sciences, Beijing 100049, China}
\author{Lan Chen}
\affiliation{Institute of Physics, Chinese Academy of Sciences, Beijing 100190, China}
\affiliation{School of Physical Sciences, University of Chinese Academy of Sciences, Beijing 100049, China}
\affiliation{Songshan Lake Materials Laboratory, Dongguan, Guangdong 523808, China}
\author{Kehui Wu}\thanks{khwu@iphy.ac.cn}
\affiliation{Institute of Physics, Chinese Academy of Sciences, Beijing 100190, China}
\affiliation{School of Physical Sciences, University of Chinese Academy of Sciences, Beijing 100049, China}
\affiliation{Songshan Lake Materials Laboratory, Dongguan, Guangdong 523808, China}
\author{Yugui Yao}\thanks{ygyao@bit.edu.cn}
\affiliation{Key Lab of Advanced Optoelectronic Quantum Architecture and Measurement (MOE), School of Physics, Beijing Institute of Technology, Beijing 100081, China}
\author{Baojie Feng}\thanks{bjfeng@iphy.ac.cn}
\affiliation{Institute of Physics, Chinese Academy of Sciences, Beijing 100190, China}
\affiliation{School of Physical Sciences, University of Chinese Academy of Sciences, Beijing 100049, China}

\date{\today}

\begin{abstract}
{Monolayer AlB$_2$ is composed of two atomic layers: honeycomb borophene and triangular aluminum. In contrast with the bulk phase, monolayer AlB$_2$ is predicted to be a superconductor with a high critical temperature. Here, we demonstrate that monolayer AlB$_2$ can be synthesized on Al(111) via molecular beam epitaxy. Our theoretical calculations revealed that the monolayer AlB$_2$ hosts several Dirac cones along the $\Gamma$--M and $\Gamma$--K directions; these Dirac cones are protected by crystal symmetries and are thus resistant to external perturbations. The extraordinary electronic structure of the monolayer AlB$_2$ was confirmed via angle-resolved photoemission spectroscopy measurements. These results are likely to stimulate further research interest to explore the exotic properties arising from the interplay of Dirac fermions and superconductivity in two-dimensional materials.}
\end{abstract}

\maketitle

The discovery of the high-temperature superconductor MgB$_2$ ($T_c$ $\approx$ 39 K) has stimulated significant research interest in the AlB$_2$-family of materials \cite{KangWN2001,NagamatsuJ2001}. In MgB$_2$, the $\sigma$-bonding boron orbitals couple strongly with the in-plane B-B stretching phonon modes \cite{BohnenKP2001,YildirimT2001,KongY2001,ChoiHJ2002}, which is crucial for the occurrence of high-temperature superconductivity. However, in AlB$_2$, an isostructural compound of MgB$_2$, the boron $\sigma$ state is located far below the Fermi level and lacks effective coupling with phonons in the boron layer \cite{BohnenKP2001,MedvedevaNI2001}. Therefore, no experimental evidence for AlB$_2$ superconductivity has been reported to date. Recently, the desire to miniaturize quantum devices has driven significant research interest in two-dimensional (2D) materials \cite{CastroNeto2009,XuM2013}. In the 2D limit, monolayer AlB$_2$ has been predicted to be a superconductor with intriguing multigap character \cite{GaoM2019,ZhaoY2019}, which is in stark contrast with the non-superconducting properties of bulk AlB$_2$. In addition, bulk AlB$_2$ has been found to host Dirac nodal lines \cite{TakaneD2018}, which indicates the possible existence of topological band structures in monolayer AlB$_2$. However, the synthesis of monolayer AlB$_2$ has remained a challenge to date and little is known about the topological properties of monolayer AlB$_2$.

Recently, various synthetic 2D materials have been realized via molecular beam epitaxy (MBE), including silicene \cite{VogtP2012,FengB2012}, stanene \cite{ZhuF2015,DengJ2018}, and borophene \cite{MannixAJ2015,FengB2016}. In particular, honeycomb borophene, an important constituent of monolayer AlB$_2$, has been realized on Al(111) \cite{LiW2018}. Notably, the topmost atomic layer of Al(111) has a flat triangular lattice that can constitute monolayer AlB$_2$ with honeycomb borophene. However, in the previously proposed structure model, the lattice of borophene was compressed to fit the lattice constant of Al(111) \cite{LiW2018,ZhuL2019,GaoM2019}, and thus the topmost triangular Al lattice was inseparable from the underlying Al(111) substrate. This results in strong hybridization of the electronic structure of AlB$_2$ with the substrate.

In this work, however, our combined low-energy electron diffraction (LEED) and scanning tunneling microscopy (STM) measurements show that the lattice constant of the surface AlB$_2$ layer is slightly larger than that of Al(111) (Fig. 1(a)-1(d)), which indicates relatively weak coupling between AlB$_2$ and Al(111). We also studied the electronic structures and topological properties of monolayer AlB$_2$ via angle-resolved photoemission spectroscopy (ARPES) and first-principles calculations. Several symmetry-protected Dirac cones were observed in a freestanding AlB$_2$ monolayer, and most of them were preserved on Al(111). Moreover, some of the Dirac bands cross the Fermi level and may contribute to electron-phonon coupling. Therefore, the realization of monolayer AlB$_2$ provides an ideal platform to study the exotic properties that arise from the coexistence of Dirac fermions and superconductivity.

The sample preparation, transfer, and measurements were all performed in ultrahigh vacuum systems with a base pressure lower than 1.0$\times$10$^{-8}$ Pa. A clean Al(111) substrate was prepared via repeated sputtering and annealing cycles. Pure boron (99.9999\%) was evaporated onto Al(111) using an e-beam evaporator. The Al(111) substrate was held at a temperature of ~500 K during growth. The STM  experiments were performed in a home-built low-temperature STM-MBE system and the data was acquired at 78 K. The LEED and ARPES measurements were performed at the BL-1 \cite{IwasawaH2017} and BL-9A of the Hiroshima synchrotron radiation center. The energy resolution of the ARPES measurements was $\sim$15 meV; the temperature of the sample during both the ARPES and LEED measurements was maintained at 30 K.

First-principles calculations were performed using the Vienna ab initio simulation package (VASP) \cite{KresseJ1993} based on the generalized gradient approximation (GGA) \cite{Blochl1994} in the Perdew-Burke-Ernzerhof (PBE) functional \cite{PerdewJP1997} and the projector augmented-wave (PAW) pseudopotential \cite{MonkhorstHJ1976}. The energy cutoff was set to 400 eV for the plane-wave basis and the Brillouin Zone was sampled using a $\Gamma$ centered Monkhorst-Pack grid \cite{PerdewJP1992} (18$\times$18$\times$1). The vacuum space was set to be larger than 20 {\AA}. All the atomic positions and lattice parameters were fully relaxed before further calculations, and the maximum force allowed on each atom was less than 0.01 eV {\AA}$^{-1}$. The numerical convergence accuracy of the total energy was 1$\times$10$^{-6}$ eV per cell. Spin-orbit coupling (SOC) effects were neglected in all our calculations. The first-principles phonon calculations were implemented in VASP and PHONOPY \cite{TogoA2015} within the framework of the density functional perturbation theory (DFPT) \cite{GonzeX1997,GiannozziP1991}.

\begin{figure}[t]
\centering
\includegraphics[width=8 cm]{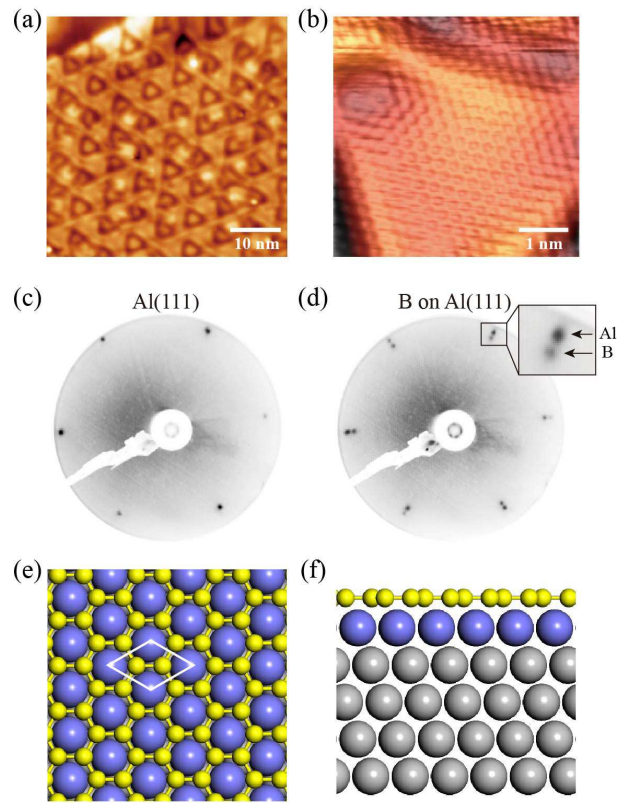}
\caption{(a) An STM image of boron on Al(111) showing the triangular corrugations. (b) Magnified STM image showing the honeycomb lattice of boron. (c) and (d) LEED patterns of Al(111) and B/Al(111), respectively. (e) and (f) Top and side views of the structure model of B/Al(111). White rhombus indicates a unit cell of honeycomb borophene or monolayer AlB$_2$. The boron and topmost Al atoms are indicated by yellow and blue balls, respectively. The underlying Al atoms are indicated by grey balls.}
\end{figure}

The growth of boron on Al(111) leads to the formation of an ordered structure with triangular corrugations, as shown in Fig. 1(a). The period of the triangular corrugation is $\sim$7 nm. From the high-resolution STM image in Fig. 1(b), a honeycomb-like structure can be observed with a lattice constant of $\sim$3.0 {\AA}, which indicates the formation of the honeycomb borophene. These results agree well with previous reports \cite{LiW2018}. LEED measurements were performed to study the atomic structure of this system. Figure 1(c) and 1(d) show the LEED patterns of pristine Al(111) and B/Al(111), respectively. It was found that the lattice constant of the surface structure was slightly larger than that of Al(111), as shown in the inset of Fig. 1(d). Based on the LEED pattern, the lattice constant of the surface structure was estimated to be 2.98 \AA, which was in qualitative agreement with the STM results. Because of the different lattice constants of the surface structure and the underlying substrate, moir\'{e} patterns form because of the lattice mismatch. A simple analysis shows that the 25$\times$25 superstructure of Al(111) ($a\rm_{Al}$=2.86 \AA) corresponds to the 24$\times$24 superstructure of the surface layer ($a\rm_s$=2.98 \AA). The period of the superstructure is $\sim$7.15 nm, which is in agreement with the period of the triangular corrugations ($\sim$7 nm). Therefore, our results confirmed that the triangular corrugations originate from the moir\'{e} patterns of the system.

\begin{figure}[htb]
\centering
\includegraphics[width=8.5cm]{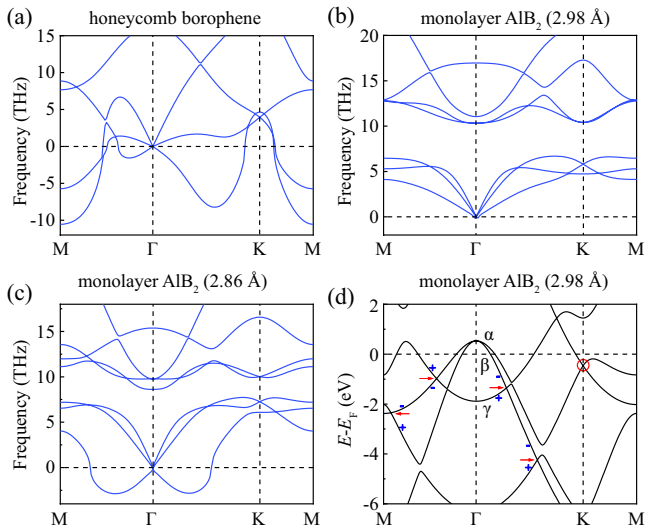}
\caption{(a) Calculated phonon spectrum of the honeycomb borophene. (b,c) Calculated phonon spectrum of the monolayer AlB$_2$ with lattice constants of 2.98 {\AA} and 2.86 {\AA}, respectively. The optimized lattice constant of the freestanding AlB$_2$ monolayer is 2.98 {\AA}. (d) Calculated band structure of the monolayer AlB$_2$. The three characteristic bands are indicated by $\alpha$, $\beta$, and $\gamma$. Red arrows indicate the Dirac cones protected by mirror symmetry along the high-symmetry lines. The ``+'' and ``-'' signs (in blue) along $\Gamma$--M and $\Gamma$-K are the mirror eigenvalues of M$_{{\Gamma}\rm{M}}$ and M$_{{\Gamma}\rm{K}}$, respectively. The red circle indicates the Dirac cone derived from the $p_z$ orbitals of boron.}
\end{figure}

There are two possibilities for the atomic structure of the surface layer: (1) only honeycomb borophene; (2) two atomic layers that contain the topmost borophene and a triangular Al, {\it i.e.}, monolayer AlB$_2$. First-principles calculations were performed to confirm the correct structure model. The optimized lattice constants of the freestanding borophene and AlB$_2$ are 2.92 {\AA} and 2.98 {\AA}, respectively. The calculated lattice constants of the freestanding AlB$_2$ monolayer were in excellent agreement with our experimental values. Figure 2(a) and 2(b) show the phonon spectrum of the freestanding honeycomb borophene and monolayer AlB$_2$. A significant imaginary frequency component can be observed for the honeycomb borophene, while no imaginary frequencies were observed for the monolayer AlB$_2$, which indicates that the monolayer AlB$_2$ is more stable than the honeycomb borophene. In addition, the monolayer AlB$_2$ will become unstable if the lattice constant is reduced to fit Al(111) ($a\rm_{Al}$=2.86 {\AA}), as shown in Fig. 2(c). Therefore, for the B/Al(111) system, we can conclude that the monolayer AlB$_2$ as a whole has a larger lattice constant than Al(111). The lattice mismatch and appearance of moir\'{e} patterns indicate a relatively weak interaction of the monolayer AlB$_2$ with the Al(111) substrate.

After establishing the synthesis of the AlB$_2$ monolayer, we move on to studying its electronic structure. Figure 2(d) shows the calculated band structure of freestanding AlB$_2$, which is in agreement with recent calculation results \cite{ZhaoY2019}. In proximity to the $\Gamma$ point, there are several bands that cross the Fermi level: $\alpha$, $\beta$, and $\gamma$. Interestingly, these bands host two Dirac cones along the $\Gamma$--K and $\Gamma$--M directions, respectively, as indicated by the red arrows in Fig. 2(d). The mirror eigenvalues of these bands are indicated by the ``+'' and ``-'' signs. The crossing bands of these Dirac cones have opposite eigenvalues, which indicates that these Dirac cones are protected by the mirror reflection symmetry: the $\Gamma$--K--$k_z$ plane and $\Gamma$--M--$k_z$ plane, respectively. Another Dirac cone is centered at the K point, as indicated by the red circle in Fig. 2(d). This Dirac cone derives from the $p_z$ orbitals of boron \cite{SM}, analogous to the Dirac cone of the honeycomb lattice. Therefore, the Dirac cone at the K point originates from the honeycomb borophene and survives in the monolayer AlB$_2$ despite the inclusion of a hexagonal Al layer.

According to the previous calculations, coupling between the boron $\sigma$ bands ({\it i.e.}, $\alpha$, and $\beta$ bands in our work) and the in-plane phonon modes gives rise to superconductivity in the AlB$_2$ monolayer \cite{GaoM2019,ZhaoY2019}. From our calculation results, some of the Dirac bands originate from the $\alpha$ and $\beta$ bands and cross the Fermi level. Therefore, the AlB$_2$ monolayer may have exotic properties that arise from the interplay of Dirac fermions and superconductivity.

\begin{figure}[t]
\centering
\includegraphics[width=8.5cm]{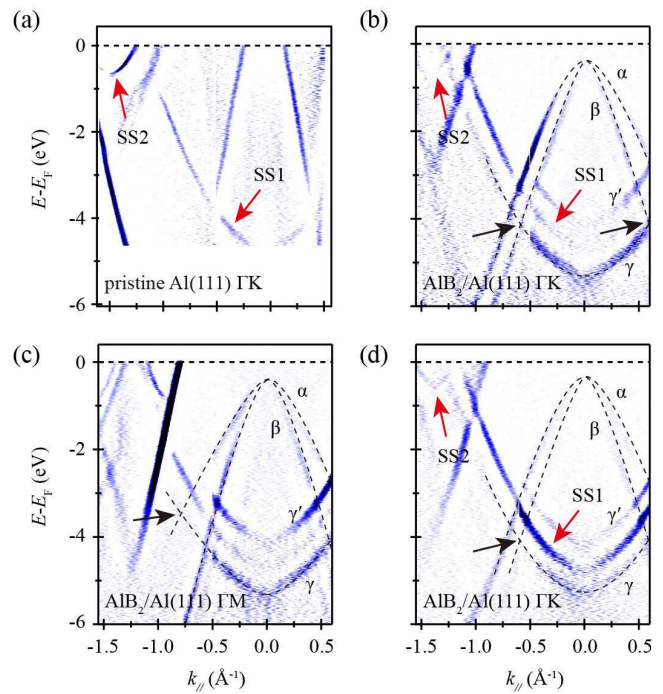}
\caption{(a) ARPES second derivative image of pristine Al(111) measured with 25-eV photons. (b) and (c) ARPES second derivative images of AlB$_2$/Al(111) along the $\Gamma$--K and $\Gamma$--M directions, respectively. $\alpha$, $\beta$, and $\gamma$ indicate the three characteristic bands of freestanding AlB$_2$. The Dirac points are indicated by black arrows. The incident photon energy is 35 eV. (d) ARPES second derivative image of AlB$_2$/Al(111) along the $\Gamma$--K direction measured with 40-eV photons. Red arrows indicate the surface states of Al(111); black arrows indicate the Dirac points of the monolayer AlB$_2$. The black dashed lines are guides for the eye. }
\end{figure}

ARPES measurements were performed to verify the intriguing electronic structures of the monolayer AlB$_2$ and the results are shown in Fig. 3. The $\alpha$, $\beta$, and $\gamma$ bands near the $\Gamma$ point of freestanding AlB$_2$ can be clearly observed in the ARPES results. In particular, the Dirac cones survive without any obvious gap opening, as indicated by the black arrows. The persistence of these bands on Al(111) indicates a weak interaction between AlB$_2$ and Al(111). There was no discernable $k_z$ dispersion on changing the incident photon energy, which agrees with the 2D character of these bands. Furthermore, an additional $\gamma^{\prime}$ band was observed, which was located 0.5 eV above the $\gamma$ band. This band originates from the hybridization of AlB$_2$ with Al(111) and will be discussed later. It should be noted that two electron-like bands were observed at the $\bar{\Gamma}$ and $\bar{K}$ points of Al(111), as indicated by the red arrows. These bands originate from the surface states of Al(111) \cite{KevanSD1985} because the coverage of AlB$_2$ was less than one monolayer. The observation of the Al(111) surface state indicates the high order and cleanliness of the sample surface.

Next, the origin of the $\gamma^{\prime}$ band is discussed. To this end, we performed first-principles calculations including the Al(111) substrate. Because of the large unit cell of the moir\'{e} pattern ($\sim$7 nm), calculating the supercell is difficult. To simplify the calculations, the lattice constant of AlB$_2$ was compressed from 2.98 {\AA} to 2.86 {\AA} to accommodate the hexagonal lattice of Al(111). The calculated band structure is shown in Fig. 4(a). It is clear that the $\alpha$ and $\beta$ bands are preserved on Al(111). The $\gamma$ band is blurred by the quantum well states arising from the finite thickness of the slab in the calculations. To distinguish the $\gamma$ band, the distance between the AlB$_2$ layer and Al(111) was increased to weaken the interaction of AlB$_2$ and the substrate. This is reasonable because the simplification in our calculation inevitably increased the coupling between AlB$_2$ and Al(111). Figures 4(b)-4(d) show the calculated band structures with different additional separations: 0.8 {\AA}, 1.4 {\AA}, and 1.8 {\AA}. We find that the $\gamma$ and $\gamma^{\prime}$ bands become more prominent with increasing separation, as highlighted by the red ellipses. When the separation was further increased, the intensity of the $\gamma^{\prime}$ band gradually decreased on the surface AlB$_2$ layer and finally disappeared. In contrast, the intensity of the $\gamma^{\prime}$ band gradually increased on the Al(111) substrate \cite{SM}. These results indicate that the $\gamma^{\prime}$ band is a hybridized state with a strong bulk character.

\begin{figure}[t]
\centering
\includegraphics[width=8.5cm]{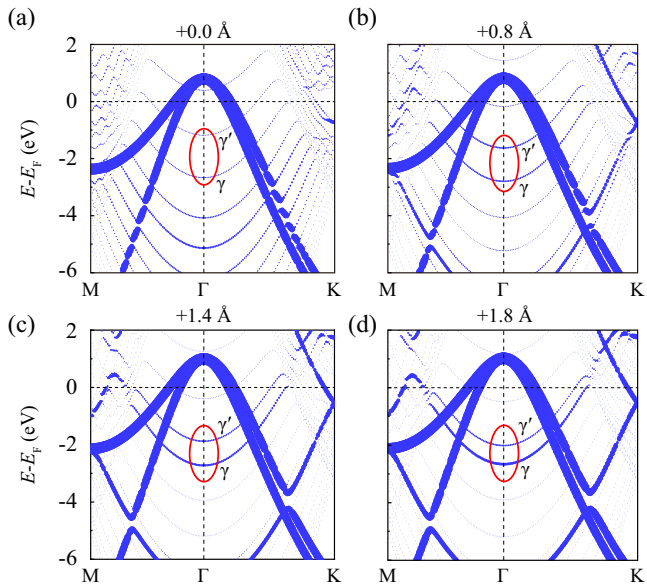}
\caption{Calculated band structures of AlB$_2$/Al(111) with different additional separations between AlB$_2$ and Al(111): 0 \AA, 0.8 \AA, 1.4 \AA, and 1.8 \AA. Red ellipses indicate the $\gamma$ and $\gamma^{\prime}$ bands. The thickness of the lines correspond to the spectral weight of the bands.}
\end{figure}

Our results support the formation of monolayer AlB$_2$ on Al(111), and more importantly, the band hybridization of AlB$_2$ and the Al(111) substrate is relatively weak. Therefore, most of the bands of the freestanding AlB$_2$ survive on Al(111). Monolayer AlB$_2$ has been predicted to be a superconductor with intriguing multigap character, hence, its successful synthesis in this work and the discovery of Dirac cones provides an ideal platform for studying the interplay of Dirac fermions and Bogoliugov quasiparticles in the 2D limit. It should be noted that the Al(111) substrate is also a superconductor ($T_c$ $\approx$ 1.2 K), which could possibly ensure the persistence of superconductivity in the AlB$_2$/Al(111) system.

\begin{acknowledgments}
This work was supported by the Ministry of Science and Tachnology of China (Grant No. 2018YFE0202700, No. 2016YFA0300904 No. 2016YFA0300600), the National Natural Science Foundation of China (Grants No. 11974391, No. 11825405, No. 1192780039, No. 11761141013, and No. 11734003), the Beijing Natural Science Foundation (Grant No. Z180007), and the Strategic Priority Research Program of Chinese Academy of Sciences (Grant No. XDB30000000). The ARPES measurements were performed with the approval of the Proposal Assessing Committee of Hiroshima Synchrotron Radiation Center (Proposal Numbers: 19AG005 and 19AG058).

\end{acknowledgments}

\end{document}